\documentclass[journal=accacs,manuscript=article,layout=two-column]{achemso}

\makeatletter
\let\citation\@gobble
\let\@citex\@gobbletwo
\makeatother

\usepackage{chemformula} 
\usepackage[T1]{fontenc} 
\usepackage{float}
\usepackage{amsmath}
\usepackage{amssymb}
\usepackage{amsthm}
\usepackage{siunitx}
\usepackage{caption} 

\SectionNumbersOn

\author{Zhen Zhu}
\author{Shan Gao}
\email{gaoshan@nbu.edu.cn}
\author{Jing Zhang}
\author{Xuxin Kang}
\affiliation[Ningbo University]
{School of Physical Science and Technology, Ningbo University, Ningbo 315211, China}
\author{Shunfang Li}
\affiliation[Zhengzhou University]
{Key Laboratory of Material Physics, Ministry of Education, School of Physics, Zhengzhou University, Zhengzhou 450001, China}
\alsoaffiliation[Second University]
{Institute of Quantum Materials and Physics, Henan Academy of Sciences, Zhengzhou 450046, China}
\author{Xiangmei Duan}
\email{duanxiangmei@nbu.edu.cn}
\affiliation[Ningbo University]
{School of Physical Science and Technology, Ningbo University, Ningbo 315211, China}

\title{Identifying the Catalytic Descriptor of Single-Atom Catalysts in Nitrate Reduction Reaction: An Interpretable Machine-Learning Method}

\abbreviations{IR,NMR,UV}

\begin{document}

\begin{tocentry}
    \includegraphics[width=0.75\textwidth, height=5cm, keepaspectratio=true]{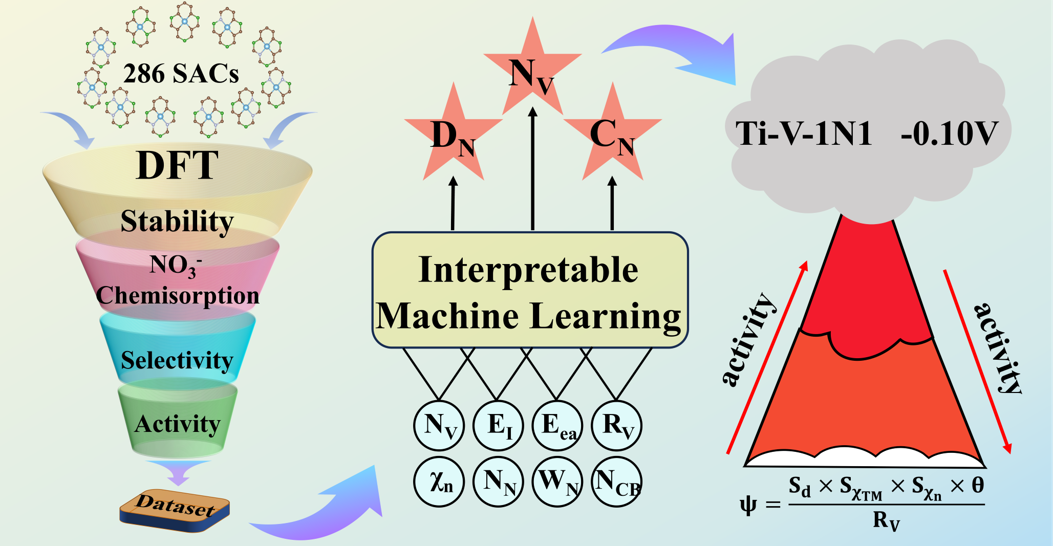}
    
\end{tocentry}

\begin{abstract}
  Elucidating the catalytic descriptor that accurately characterizes the structure-activity relationships of typical catalysts for various important heterogeneous catalytic reactions is pivotal for designing high-efficient catalytic systems. Here, an interpretable machine learning technique was employed to identify the key determinants governing the nitrate reduction reaction ($\rm NO_3RR$) performance across 286 single-atom catalysts (SACs) with the active sites anchored on double-vacancy $\rm BC_3$ monolayers. Through Shapley Additive Explanations (SHAP) analysis with reliable predictive accuracy, we quantitatively demonstrated that, favorable $\rm NO_3RR$ activity stems from a delicate balance among three critical factors: low $\rm N_V$, moderate $\rm D_N$, and specific doping patterns. Building upon these insights, we established a descriptor ($\psi$) that integrates the intrinsic catalytic properties and the intermediate O-N-H angle ($\theta$), effectively capturing the underlying structure-activity relationship. Guided by this, we further identified 16 promising catalysts with predicted low limiting potential ($U_{\rm L}$). Importantly, these catalysts are composed of cost-effective non-precious metal elements and are predicted to surpass most reported catalysts, with the best-performing Ti-V-1N1 is predicted to have an ultra-low $U_{\rm L}$ of $-0.10$ V.
\end{abstract}

\section{Introduction}
The dual challenges of nitrogen-related environmental remediation and sustainable ammonia synthesis have become critical focal points in modern catalysis research. Excessive nitrate ($\rm NO_3^-$) contamination in aquatic ecosystems, primarily from agricultural runoff and industrial effluents, poses severe threats to biodiversity and public health,\textsuperscript{\cite{popova2023evaluating,xu2022electrocatalytic}} while its chemical stability creates substantial thermodynamic barriers for conventional remediation strategies.\textsuperscript{\cite{clark2019mechanistic,wei2020mild}} Concurrently, the global demand for ammonia ($\rm NH_3$) -- an indispensable precursor for fertilizers (80$\%$ of global production), pharmaceuticals, and advanced materials -- continues to escalate. This demand is currently met through the century-old Haber-Bosch process, which, however, is an energy-intensive technology consuming 1-2$\%$ of global energy output and responsible for 1.5$\%$ of anthropogenic $\rm CO_2$ emissions.\textsuperscript{\cite{zheng2022green,scholes2021nitrate}} Accordingly, electrochemical nitrate reduction reaction ($\rm NO_{3}RR$) presents an innovative paradigm to address both challenges simultaneously, converting environmental pollutants into value-added $\rm NH_3$ through renewable electricity.\textsuperscript{\cite{wang2023renewable, wang2022iron, theerthagiri2022electrocatalytic}}

Compared to alternative nitrogen sources ($\rm N_2$: 941 kJ/mol, NO: 631 kJ/mol), nitrate demonstrates superior catalytic accessibility due to its lower N-O bond dissociation energy ($\rm 204\ kJ/mol$) and high aqueous solubility ($\rm \textgreater 1\ M$ at \SI{25}{\degreeCelsius}), enabling it thermodynamically favorable conversion pathways.\textsuperscript{\cite{jeon2024recent,fernandez2020nitrate}} While noble metal catalysts (e.g., Ru, Pt) have shown promising $\rm NO_3RR$ activity,\textsuperscript{\cite{li2020efficient}} their practical implementation is hindered by prohibitive costs and limited durability. In recent decades, single-atom catalysts (SACs) have emerged as revolutionary materials in heterogeneous catalysis, offering atomic-level efficiency, exceptional stability, and precisely tunable active sites through coordination engineering.\textsuperscript{\cite{sun2024advancing,chi2023structural,liu2023regulating}} Two primary design strategies dominate SAC optimization: (1) Transition metal (TM) selection to regulate $d\mbox{-}{\rm band}$ electronic structure, as evidenced by Ti/Zr SACs on g-CN monolayers achieving record-low overpotentials;\textsuperscript{\cite{niu2021theoretical}} (2) Strategic heteroatom doping (N, P, S) in primary/secondary coordination spheres, exemplified by S/P-coordinated SACs demonstrating enhanced intermediate adsorption energetics,\textsuperscript{\cite{zang2023activating}} and TM-based SACs optimized through synergistic ligand engineering.\textsuperscript{\cite{shin2023dft,xu2018retracted,xia2023ultrastable}} These approaches can induce divergent reaction pathways and improved product selectivity.

Despite these advances, fundamental challenges persist in establishing quantitative structure-activity relationships (QSARs) for SACs. The inherent complexity of $\rm NO_3RR$ mechanisms, involving eight-electron transfers and multiple intermediates ($\rm ^*NO_3$ $\rightarrow$ $\rm ^*NO_2$ $\rightarrow$ $\rm ^*NO$ $\rightarrow$ $\rm ^*NH_3$), creates multidimensional parameter spaces that challenge conventional analysis.\textsuperscript{\cite{sun2024interpretable}} While high-throughput density function theory (DFT) screening has accelerated catalyst discovery, its reliance on simplified descriptor models (e.g., $d\mbox{-}{\rm band}$ center, work function) often overlooks critical coordination environment effects.\textsuperscript{\cite{liao2022density, huang2022single, chen2021high, wang2022high, ou2022silico,sun2024interpretable,li2016interplay}} Machine learning (ML) methodologies offer transformative potential in decoding these complex correlations.\textsuperscript{\cite{wu2023target, zhang2023accelerated,weng2020simple,zheng2023predicting}} Particularly, interpretable machine learning (IML) techniques like Shapley Additive Explanations (SHAP) enable quantitative feature importance analysis while maintaining predictive accuracy, particularly valuable for SAC systems where metal-center properties, coordination environments, and substrate interactions create high-dimensional design spaces.\textsuperscript{\cite{nadernezhad2022machine,lu2022unraveling,liu2025screening,baryannis2019predicting,ibarguren2022pctbagging}} 

Here, we present a synergistic computational framework combining IML with DFT to establish fundamental design principles for $\rm NO_3RR$ SACs. Through systematic investigation of 286 distinct SAC configurations, comprising 26 TMs anchored on $\rm BC_3$ divacancy substrates, we first identified 56 promising candidates via high-throughput DFT screening. To address the inherent data imbalance (active vs. inactive catalysts), we implement an XGBoost model with synthetic minority over-sampling method, achieving exceptional performance. Coupled with SHAP analysis, we revealed three dominant performance determinants: (i) The number of valence electrons of reactive TM single atom ($\rm N_V$), (ii) doping concentration of nitrogen ($\rm D_N$), and (iii) coordination configuration of nitrogen ($\rm C_N$). Based on these features and the characteristic O-N-H bond angle ($\theta$) of key intermediates, we further develop a multidimensional descriptor ($\psi$) that exhibits a volcano-shaped relationship with the limiting potential ($U_{\rm L}$) across the catalyst space, highlighting the critical role of the TM center and its coordination environment. Practical application of this descriptor further identifies 16 non-precious metal SACs with exceptional performance ($U_{\rm L}$ \textless \  $-$0.36 V), including the Ti-V-1N1 configuration ($U_{\rm L}$ = $-$0.10 V), which is predicted to exhibit state-of-the-art activity and confirming the predictive power of the present work. Crucially, all identified catalysts utilize earth-abundant elements while maintaining superior efficiency, establishing a new paradigm for sustainable electrochemical nitrate remediation.

\section{Computational Methods}

All spin-polarized DFT calculations were performed using the Vienna $ab$ $initio$ Simulation Package (VASP).\textsuperscript{\cite{kresse1996efficient, kresse1999ultrasoft}} The electron correlation interactions were described by the generalized gradient approximation (GGA) with the Perdew-Burke-Ernzerhof (PBE) exchange-correlation functional.\textsuperscript{\cite{kresse1999ultrasoft,perdew1996generalized}} The geometric optimization and electronic structure calculations were carried out using 4 $\times$ 4 $\times$ 1 and 9 $\times$ 9 $\times$ 1 Monkhorst-Pack grids for Brillouin zone sampling, respectively. Van der Waals interactions were corrected with Grimme's DFT-D3 method.\textsuperscript{\cite{goerigk2014dft}} A 20 \AA \ vacuum layer along the $Z$ direction was introduced to avoid periodic interlayer interactions. The kinetic cutoff energy was chosen to be 520 eV, with force and energy calculation accuracy set as $10^{-2}$ eV \AA$\rm ^{-1}$ and $10^{-5}$ eV, respectively. All structures were successfully relaxed to meet the specified criteria. Thermodynamic stability was evaluated through $ab$ $initio$ molecular dynamics (AIMD) simulations,\textsuperscript{\cite{martyna1992nose}} conducted at 500 K for a duration of 6 ps.

The binding energy ($E_{\rm b}$) of TM atoms anchored at the defect site is defined as:
\begin{equation}
    E_{\rm b} = E_{\rm total} - E_{\rm TM} - E_{\rm defect}
\end{equation}
where $E_{\rm total}$ and $E_{\rm TM}$ represent the energy of the system after anchoring the TM atom and a single TM atom, respectively.

In order to avoid the direct calculation of the energy of charged $\rm NO_3^-$, gaseous $\rm HNO_3$ was chosen as a reference.\textsuperscript{\cite{liu2019activity}} The Gibbs free energy change of $\rm NO_3^-$ adsorption ($\Delta G_{\rm ^*NO_3}$) is expressed as:
\begin{equation}
    \begin{split}
            \Delta G_{\rm ^*NO_3} = G_{\rm ^*NO_3} - G_{\rm ^*} - G_{\rm HNO_3(g)} \\ +\ 0.5G_{\rm H_2(g)} + \Delta G_{\rm correct}
    \end{split}
\end{equation}
\begin{equation}
    \Delta G_{\rm correct} = - \Delta G_{\rm S1} - \Delta G_{\rm S2}
\end{equation}
where $G_{\rm ^*NO_3}$, $G_{\rm ^*}$, $G_{\rm HNO_3(g)}$ and $G_{\rm H_2(g)}$ are the Gibbs free energy of SACs with and without $\rm NO_3^-$ adsorption, $\rm HNO_3$ and $\rm H_2$ molecules in the gas phase, respectively. $\Delta G_{\rm correct}$ is the free energy correction for gaseous $\rm HNO_3$ dissolution ($\Delta G_{\rm S1}$) as well as electronic processes ($\Delta G_{\rm S2}$) . Herein, the corresponding values are $\Delta G_{\rm S1}$ = $-$0.08 eV and $\Delta G_{\rm S2} = -$0.32 eV. Therefore, the value of $\Delta G_{\rm correct}$ is 0.08 eV $+$ 0.32 eV $=$ 0.40 eV.\textsuperscript{\cite{lv2021computational}}

The Gibbs free energy change ($\Delta G$) for each reaction step is calculated based on the computational hydrogen electrode (CHE) model:\textsuperscript{\cite{norskov2004origin}}
\begin{equation}
    \Delta G = \Delta E + \Delta E_{\rm ZPE} - T\Delta S
\end{equation}
where $\Delta E$, $\Delta E_{\rm ZPE}$ and $\Delta S$ represent the changes in energy, zero-point energy, and entropy of the adsorbed intermediates, and $T$ is the temperature ($T = 298.15$ K). The potential determining step (PDS) refers to the reaction step with the greatest variation in Gibbs free energy ($\Delta G_{\rm max}$). The limiting potential ($U_{\rm L}$) of the entire reaction process is calculated by $U_{\rm L}$ = $-\Delta G_{\rm max}/e$.

\section{Results and discussion}

\subsection{Construction of SAC Structures}

The construction of SAC systems is based on a two-dimensional (2D) $\rm BC_3$ monolayer, which has been successfully synthesized and extensively characterized in previous studies.\textsuperscript{\cite{yang2021capture, zinin2012phase}} This 2D material features a hexagonal lattice symmetry analogous to that of graphene. In our computational models, a 2 $\times$ 2 $\times$ 1 supercell was employed, yielding optimized lattice constants of a $=$ b $=$ 5.17 \AA. Geometric relaxation revealed characteristic bond lengths of 1.42 \AA \ for C-C and 1.56 \AA \ for C-B, consistent with previously reported theoretical values.\textsuperscript{\cite{ou2022silico}} To introduce active sites, we engineered two types of vacancy defects into the $\rm BC_3$ lattice: a C-C vacancy ($\rm V_{CC}$) and a C-B vacancy ($\rm V_{CB}$). The $E_{\rm f}$ for these vacancies are 6.73 eV and 7.78 eV, respectively, which are comparable to the energy required to form a single vacancy in graphene (7.65 eV).\textsuperscript{\cite{olsson2019adsorption}} 

Then, we systematically anchored 26 TM atoms spanning the 3\textit{d}, 4\textit{d}, and 5\textit{d} series at these vacancy sites, resulting in two primary SAC configurations: TM-$\rm V_{CC}$ and TM-$\rm V_{CB}$ (Fig.~\ref{fig1}a). Furthermore, inspired by the established structure-activity relationships in graphene-like systems,\textsuperscript{\cite{chen2021engineering, li2023unraveling, wen2022sulfur, bai2023enhanced, wang2024high}} we introduced nitrogen coordination to modulate the primary coordination environment of the $\rm TM\mbox{-}V_{CB}$ systems. This design strategy yielded nine distinct TM-N coordination motifs, denoted as TM-V-$\rm n_1Nn_2$, where $\rm n_1$ and $\rm n_2$ specify the number and spatial arrangement of nitrogen dopants, respectively (Fig.~\ref{fig1}a). In total, 286 distinct SAC structures were constructed (26 TMs $\times$ 11 coordination environments). Subsequent structural optimization confirmed the dynamic and thermodynamic stability of all configurations, with negligible lattice distortion observed.

Recent advances in nanomaterials synthesis, particularly precision techniques such as ion implantation, atomic layer deposition, thermal carbonization, and potential-controlled electrodeposition, have demonstrated the practical feasibility of fabricating graphene-like SAC substrates with atomic precision.\textsuperscript{\cite{chen2022copper, wu2021electrochemical,sun2013single}} These experimental breakthroughs not only validate the structural plausibility of our proposed $\rm BC_3$-based SACs, but also help bridge the gap between theoretical feasibility and real-world application. The excellent agreement between computational feasibility and synthetic accessibility positions $\rm BC_3$ monolayer as a promising and versatile platform for developing next-generation high-performance SAC materials.

\subsection{High-throughput Screening Strategies to Screen Qualified Catalysts}

\begin{figure*}[t]
	\centering
	\includegraphics[width=0.85\textwidth]{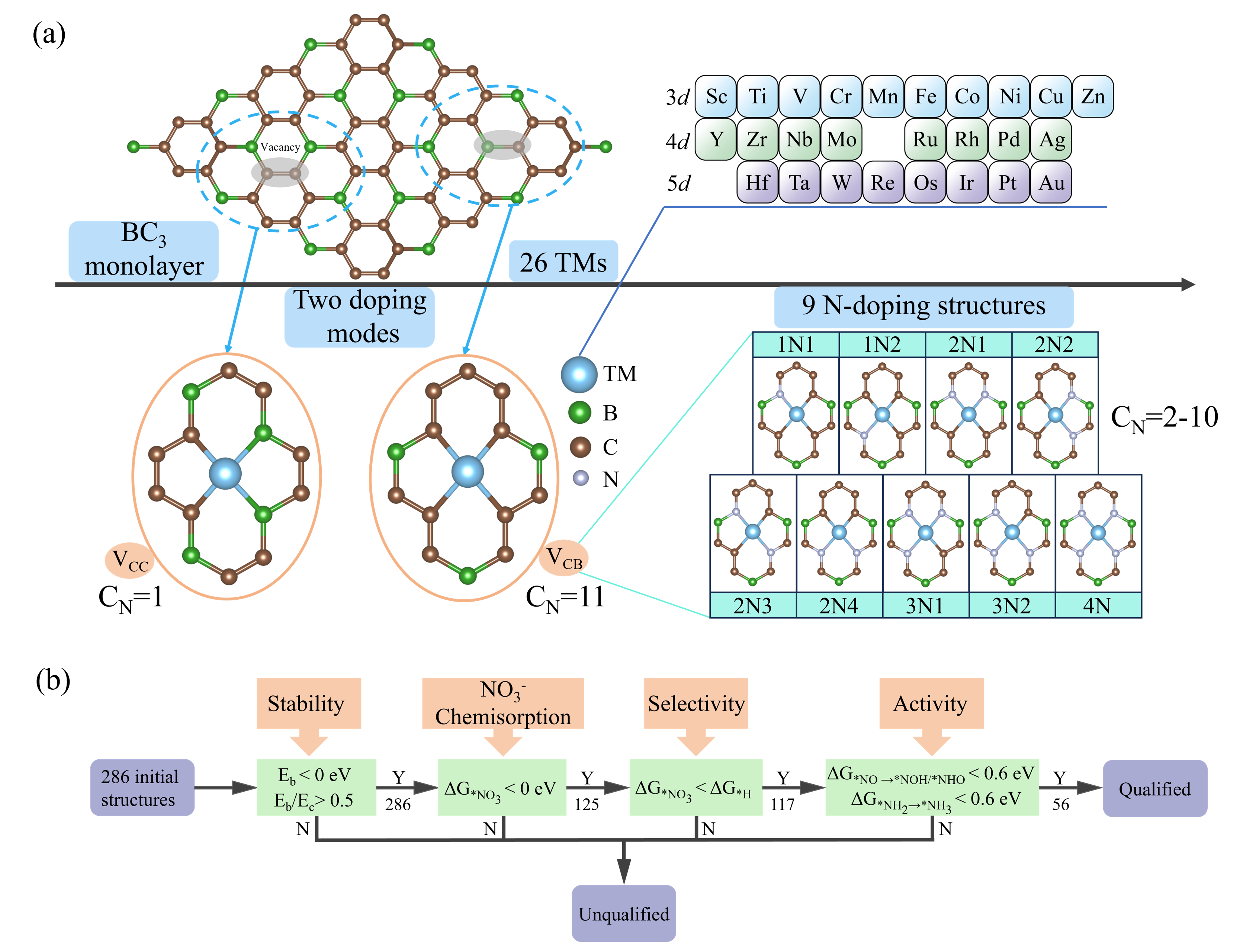}	
	\caption{(a) Schematic illustration of the catalyst space construction for $\rm TM{@}V_{CC}$ and $\rm TM{@}V_{CB}$-$\rm N_{n=0\mbox{-}4}$ configurations. Atomic species are color-coded as follows: transition metal (TM, sky blue), boron (B, green), carbon (C, brown), and nitrogen (N, silvery-white). (b) Systematic workflow of the four-stage high-throughput screening protocol employed for identifying optimal catalyst candidates.}
	\label{fig1}%
\end{figure*}

A four-step high-throughput screening strategy was implemented to effectively prioritize viable catalysts from an initial pool of 286 structural candidates, as illustrated in Fig.~\ref{fig1}b. The first screening stage evaluated the thermodynamic stability of TM atoms anchored at vacancy sites. The key criterion employed was $E_{\rm b}$, with all structures showing negative values ($E_{\rm b}$ \textless \ 0 eV), confirming the energetic favorability of TM atoms anchoring (Fig. S1a).

\renewcommand{\floatpagefraction}{.9}

The second screening stage targeted agglomeration resistance by analyzing the ratio of TM binding energy to its cohesive energy ($E_{\rm C}$). A threshold of $E_{\rm b}$/$E_{\rm C}$ \textgreater \ 0.5 was established to ensure sufficient resistance against TM clustering.\textsuperscript{\cite{liu2012growth, choi2018suppression}} Cohesive energy values were assessed using both experimentally derived ($E_{\rm C-{exp}}$) and computationally predicted ($E_{\rm C-{cal}}$) data (Fig. S1b, c), with full compliance observed across all 286 configurations. This dual-validation approach reinforced the reliability of our screening criteria from both theoretical and experimental perspectives.

The third screening stage was conducted on a randomly selected subset of 260 structures, reserving the remaining 26 systems for later predictive validation and examination. Thermodynamically favorable chemisorption of $\rm NO_3^-$ ($\Delta G_{\rm ^*NO_3}$ \textless \ 0 $\rm eV$) was verified by comparing monodentate and bidentate adsorption modes (Fig. S2). The binding configuration with lower $\Delta G_{\rm ^*NO_3}$ was selected for each structure. Nevertheless, those configurations showing weak adsorption affinity or positive adsorption free energy ($\Delta G_{\rm ^*NO_3}$ \textgreater \ 0 $\rm eV$) were excluded, reducing the number of candidates to 125.

\begin{figure*}[t]
    \centering
    \includegraphics[width=1.0\textwidth]{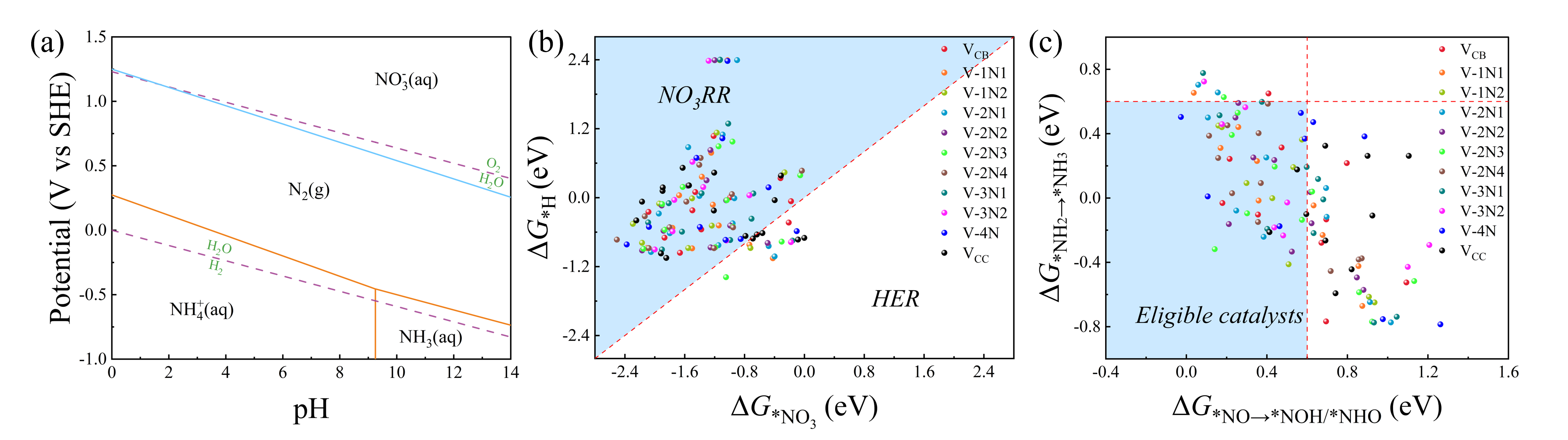}
    \caption{(a) Electrochemical stability analysis through Pourbaix diagram construction for nitrogen species. (b) Competitive adsorption profile comparison between hydrogen protons and nitrate ions ($\rm NO_3^-$) at active sites. (c) Scatter plot visualization of catalyst candidates filtered by thermodynamic criteria, specifically Gibbs free energy changes ($\Delta G $) for critical hydrogenation steps: $\rm {^*NO \rightarrow{^*NOH/^*NHO}}$ and $\rm {^*NH_2 \rightarrow{^*NH_3}}$ (threshold: \textless \ 0.6 eV.)}
    \label{fig2}
\end{figure*}

To eliminate candidates prone to the competing hydrogen evolution reaction (HER), a process that hampers $\rm NO_3RR$ selectivity, the hydrogen adsorption free energy ($\Delta G_{\rm ^*H}$) of the 125 candidates was computed. In doing this, a selectivity criterion was applied, that is, only systems that satisfied $\Delta G_{\rm ^*NO_3}$ \textless \ $\Delta G_{\rm ^*H}$ were retained. This ensured preferential nitrate adsorption over hydrogen, ultimately yielding 117 promising candidates that populate the blue region above the diagonal line in the $\Delta G_{\rm ^*NO_3}$ versus $\Delta G_{\rm ^*H}$ plot (Fig.~\ref{fig2}b).

The subsequent screening stage focused on the key intermediate $\rm ^*NO$, which lies along the experimentally validated $\rm NO_3RR$ pathway: $\rm NO_3^-$ $\rightarrow$ $\rm ^*NO_3$ $\rightarrow$ $\rm ^*NO_2$ $\rightarrow$ $\rm ^*NO$.\textsuperscript{\cite{liu2024regulating, yang2025situ, wu2023ag}} Of particular importance are the downstream hydrogenation steps, namely $\rm ^*NO$ $\rightarrow$  $\rm ^*NOH/^*NHO$ and $\rm ^*NH_2$ $\rightarrow$  $\rm ^*NH_3$. Both steps exhibit thermodynamic interdependence, often showing an inverse correlation in their Gibbs free energy changes.\textsuperscript{\cite{chen2022copper, lv2021computational, niu2021theoretical}} Moreover, given that $\rm NH_3$ readily converts to $\rm NH_4^+$ under typical pH condition (0-9) and electrochemical potentials (Fig.~\ref{fig2}a),\textsuperscript{\cite{garcia2018electrocatalytic,liu2023computational}} the desorption of the final product is generally not considered rate-limiting. While strong binding catalysts may still exhibit some thermodynamic driving force associated with $\rm NH_3$ desorption, it was not included as a primary screening criterion in this work. Instead, a dual-criterion threshold was established to evaluate the catalytic viability of remaining structures: $\Delta G_{\rm ^*NO\rightarrow^*NOH/^*NHO}$ \textless \ 0.6 eV and $\Delta G_{\rm ^*NH_2\rightarrow^*NH_3}$ \textless \ 0.6 eV. This final screening step identified 56 qualified SAC candidates (Fig.~\ref{fig2}c), each capable of energetically facilitating both critical hydrogenation transitions in the $\rm NO_3RR$ pathway.

\renewcommand{\floatpagefraction}{.9}

\begin{figure*}[t]
    \centering
    \includegraphics[width=0.7\textwidth]{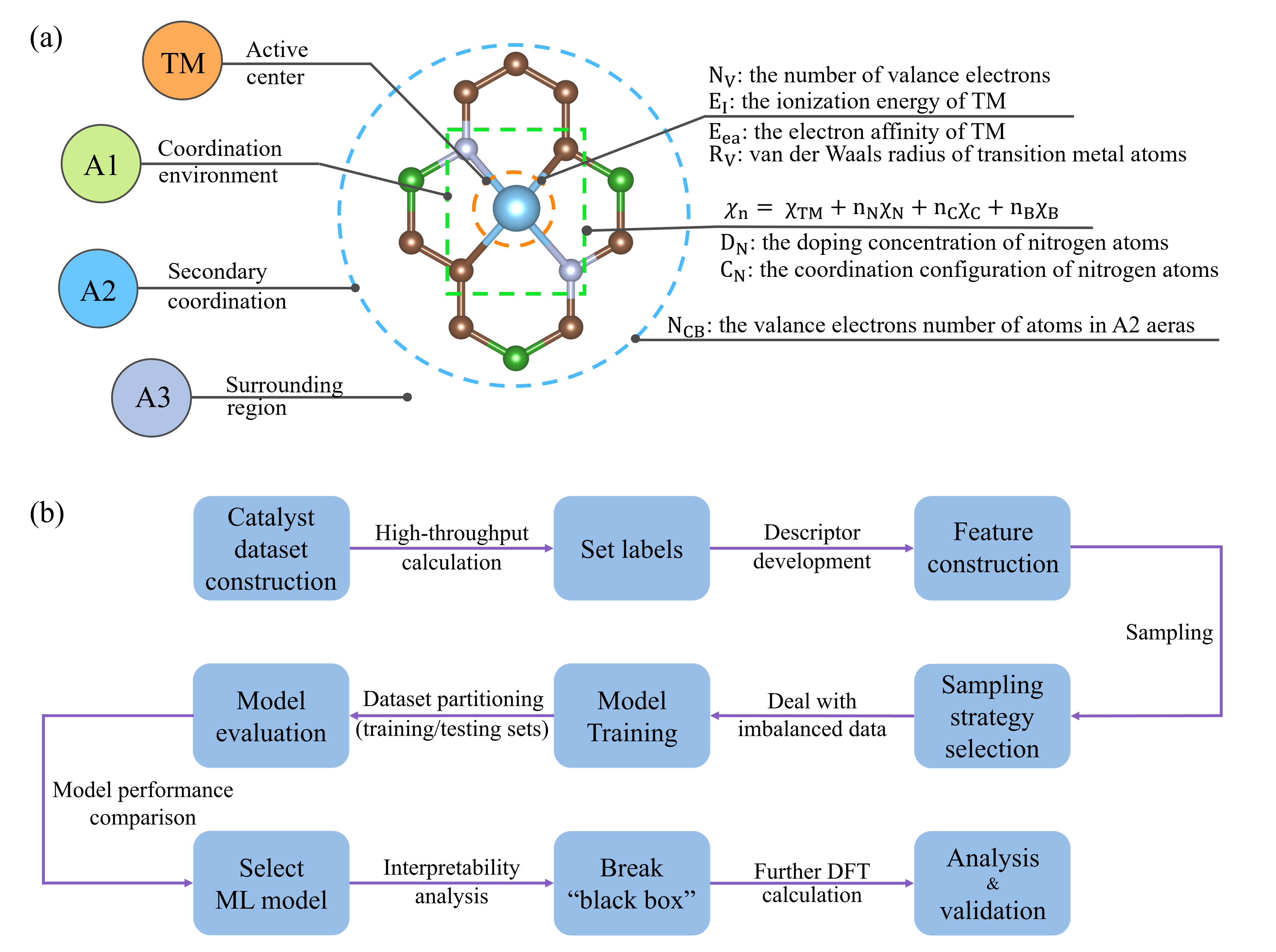}
    \caption{(a) Systematic feature engineering framework illustrating attribute characterization and corresponding active site segmentation. (b) Integrated high-throughput screening methodology coupled with comprehensive machine learning workflow architecture.}
    \label{fig3}
\end{figure*}

Building upon these 56 filtered candidates, we addressed the complexity of interpreting structure-performance relations within a dataset rich in chemical diversity and combinatorial variability. To uncover the key activity-determining features, we applied interpretable machine learning (IML) techniques. Our approach began with comprehensive feature engineering and the construction of a structured descriptor dataset. As illustrated in Fig.~\ref{fig3}a, the interfacial region surrounding the active site was hierarchically partitioned into four domains: (1) the TM active center, (2) the primary coordination shell (A1), (3) the secondary coordination environment (A2), and (4) the peripheral region (A3). For the TM region, we incorporated intrinsic elemental descriptors including valence electron count ($\rm N_V$), ionization energy ($\rm E_I$), electron affinity ($\rm E_{ea}$), and van der Waals radius ($\rm R_V$). The electronic exchange potential between the TM site and its immediate coordinating ligands (A1 domain) was quantified using the total electronegativity ($\rm \chi _n$) of atoms in this region. Furthermore, nitrogen doping characteristics, central to the electronic modulation of SACs, were captured using two descriptors: nitrogen doping concentration ($\rm D_N$) and doping configuration ($\rm C_N$). The latter was encoded numerically as follows: TM-$\rm V_{CB}$ ($\rm C_N=1$), and the nitrogen-coordinated variants TM-V-1N1 to TM-V-4N ($\rm C_N=2\mbox{-}10$), and TM-$\rm V_{CC}$ ($\rm C_N=11$) depending on nitrogen count and arrangement (Fig.~\ref{fig1}a).

To briefly account for broader electronic effects, the cumulative number of carbon and boron atoms ($\rm N_{CB}$) within the A2 region was calculated, reflecting longer-range chemical communication. In total, eight physicochemically meaningful descriptors were compiled and used as predictive input variables, while the high-throughput screening outcomes served as the supervised classification targets. This feature-based IML framework, summarized in Fig.~\ref{fig3}b, enabled the systematic elucidation of underlying structure-activity relationships and the identification of key design parameters important for $\rm NO_3RR$ catalytic performance.

\subsection{Feature Engineering and Machine Learning}

\renewcommand{\floatpagefraction}{.9}

\renewcommand{\floatpagefraction}{.9}
\begin{figure*}[t]
    \centering
    \includegraphics[width=0.8\textwidth]{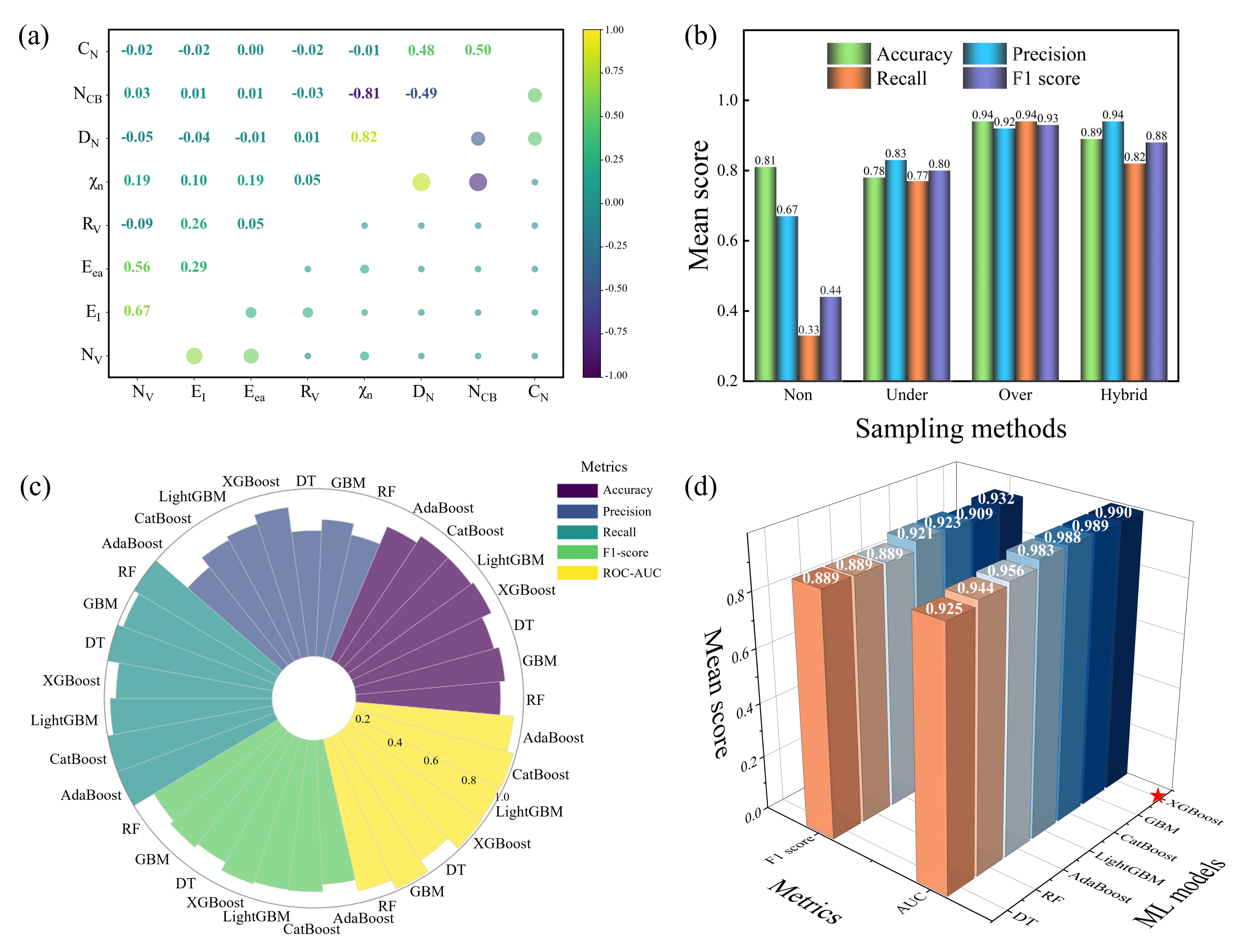}
    \caption{(a) Heat map illustrating Pearson coefficients among eight selected features. (b) Comparative performance evaluation of four sampling methodologies on XGBoost model through ten-fold cross-validation, including accuracy, precision, recall, and F1-score metrics. (c) Comprehensive model assessment of seven machine learning algorithms employing ten-fold cross-validation with over-sampling technique, presenting accuracy, precision, recall, F1-score, and AUC values. (d) Statistical distribution analysis of model performance metrics (F1-score and AUC values) across seven algorithms utilizing ten-fold cross-validation with over-sampling approach.}

    \label{fig4}
\end{figure*}

The interdependence among the eight selected features was quantitatively assessed using Pearson correlation analysis, with the resulting correlation matrix visualized in the form of a heatmap in Fig.~\ref{fig4}a. Notably strong correlations were observed between $\rm D_N$ and $\rm \chi _n$ ($\rm r=0.82$), as well as a significant inverse relationship between $\rm N_{CB}$ and $\rm \chi _n$ ($\rm r=-0.81$). To address potential predictive bias and instability due to multicollinearity, we employed an ensemble of tree-based machine learning algorithms. These include Decision Tree (DT), Random Forest (RF), Gradient Boosting Machine (GBM), Adaptive Boosting (AdaBoost), Extreme Gradient Boosting (XGBoost), Light Gradient Boosting Machine (LightGBM), and Categorical Boosting (CatBoost). Given the pronounced class imbalance, i.e., 56 qualified cases versus 204 unqualified instances, we systematically evaluated four sampling strategies: baseline non-sampling, under-sampling, over-sampling, and hybrid-sampling, and all 260 sampling protocols were implemented within an 80$\%$/20$\%$ training-test splitting scheme to ensure consistency in model evaluation.

Then, model efficacy was benchmarked through stratified ten-fold cross-validation across four evaluation axes: Accuracy, Precision, Recall, and F1-score. The comparative performance results are summarized in Fig.~\ref{fig4}b and Fig. S3. Using XGBoost as a representative case (Fig.~\ref{fig4}b), the non-sampling approach performed poorly due to unadjusted class imbalance. In contrast, over-sampling significantly improved performance across all evaluation metrics. In general, non-sampling methods consistently yielded the lowest scores, whereas both over-sampling and hybrid sampling approaches led to notable performance enhancements. Among the seven algorithms evaluated, hierarchical performance comparison revealed that over-sampling produced the most consistent evaluation metrics for CatBoost, DT, GBM, and XGBoost models (Fig. S3). Therefore, over-sampling was adopted as the standardized sampling strategy for subsequent analyses.

To further examine classification robustness, the Area Under the Receiver Operating Characteristic Curve (AUC) was incorporated as an additional evaluation criterion, complementing the established metrics. This multi-metric evaluation framework enabled a more comprehensive assessment of model performance across diverse decision thresholds. All seven models demonstrated strong classification capabilities, with AUC values consistently exceeding 0.70, as illustrated in Fig.~\ref{fig4}c.

Detailed comparative analysis is emphasized on F1-score and AUC metrics to delineate model-specific differences in predictive efficacy. Among the candidate models, XGBoost emerged as the top performer, achieving superior outcomes in both F1-score and AUC (Fig.~\ref{fig4}d). These results were further supported by ROC curve analysis and AUC distributions obtained from ten-fold cross-validation, presented in Fig. S4. Based on this thorough and rigorous methodological validation, the XGBoost model was selected as the optimal predictive framework for future high-throughput screening of catalytic activity.  

\subsection{SHAP Explainability Analysis}

By synergistically integrating high-throughput screening with ML modeling, we successfully achieved accurate classification of catalytic performance across a wide range of structural configurations. Nevertheless, the complex and nonlinear relationships that govern the transition from multiple input features to binary `qualified' or `unqualified' outcomes remain largely elusive. To unravel this complexity, we employed IML techniques to identify the key features that determine catalytic efficacy and to explore their interaction effects. This approach allowed us to demystify the `black-box' nature of ML models and enhance their transparency. In particular, we utilized SHAP for its high efficiency and fine-grained interpretability,\textsuperscript{\cite{arreche2024xai}} which enabled us to quantitatively attribute the contribution of each feature to individual predictions, thereby providing mechanistic insights into the model's decision-making process. 

\renewcommand{\floatpagefraction}{.95}
\begin{figure*}[t]
    \centering
    \includegraphics[width=0.75\textwidth]{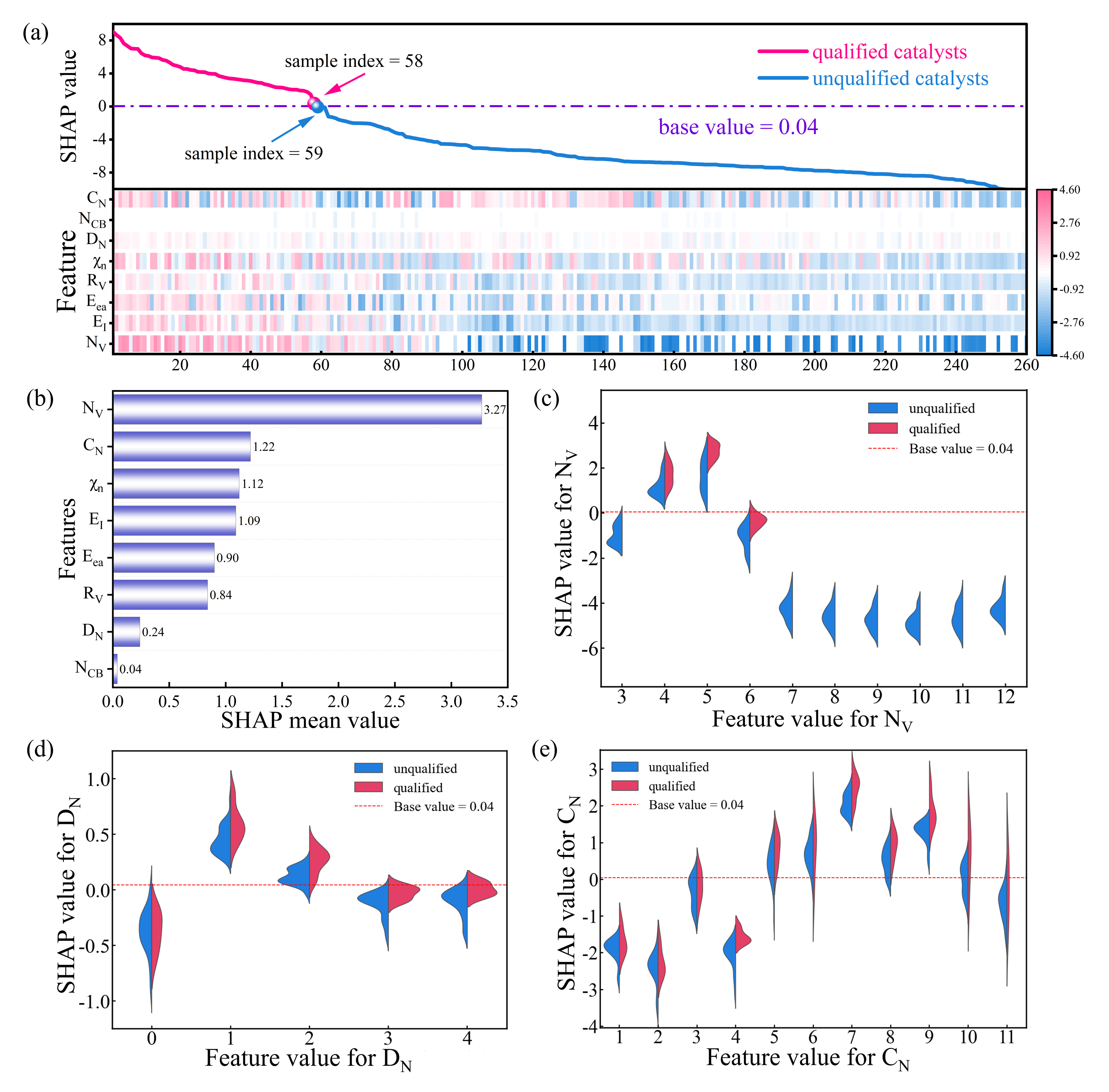}
    \caption{(a) Feature importance visualization through SHAP value heatmap analysis across 260 samples based on the XGBoost model, demonstrating classification efficacy between qualified and unqualified catalysts using a decision threshold of $0.04$. (b) Quantitative representation of feature significance through SHAP value distribution for eight critical characteristics. (c-e) Multivariate analysis of key features: (c) valence electron count: $\rm N_V$, (d) nitrogen atom concentration: $\rm D_N$ and (e) nitrogen doping configuration: $\rm C_N$, presented through composite SHAP dependence plots. Density distributions of catalyst performance are color-coded: qualified (red) and unqualified (blue) samples.}
    \label{fig5}
\end{figure*}

Fig.~\ref{fig5}a presents the SHAP contribution spectrum across all features for each sample, overlaid with cumulative SHAP scores. The SHAP decision threshold is set at 0.04, with catalysts exceeding this value classified as `qualified'. While experimental screening identified 56 qualified candidates, the SHAP-based prediction yielded 58 samples above the threshold. A closer examination revealed four misclassifications attributed to inherent model limitations: one qualified catalysts (Zr-V-2N3), was incorrectly excluded, while three unqualified catalysts (Zr-V-3N1, Mo-V-3N1, and Ta-V-2N3) were falsely included. Notably, these outliers exhibited Gibbs free energy changes of 0.57 eV, 0.68 eV, 0.65 eV, and 0.63 eV, respectively, for the critical hydrogenation steps, which lie near the screening cutoff of 0.6 eV. Despite this, the misclassification rate remains low at 1.5\% (4 out of 260), indicating that these deviations exert minimal impact on the overall interpretive reliability of the SHAP analysis. 

For global feature importance assessment, we calculated mean SHAP values per feature, as presented in Fig.~\ref{fig5}b. Here, the mean SHAP value refers to the average of the absolute SHAP values of a given feature across all samples, which quantitatively reflects the global importance of the corresponding feature. Notably, $\rm N_V$ emerged as the dominant determinant with a SHAP value of 3.27, while other features showed comparatively lower contributions (0.04 to 1.22). Supplementary SHAP summary plots (Fig. S5) provided comprehensive global interpretability though the broad distribution and discontinuous positive contributions of certain features warrant further granular analysis to fully clarify their operational modalities.

To clear up the individual contributions of key tunable parameters to catalytic competency, we conducted a systematic analysis of $\rm N_V$, $\rm D_N$, and $\rm C_N$ through SHAP dependency mapping based on the results in Fig.~\ref{fig5}a and b. The SHAP dependence plot for $\rm N_V$ (Fig.~\ref{fig5}c) illustrates distinct performance, with an $\rm N_V$ value of 4 or 5 corresponding to elevated SHAP indices indicative of enhanced catalytic efficacy. Conversely, deviations from this optimal range correlate with diminished activity.

Comparable SHAP dependency analysis of $\rm D_N$ (Fig.~\ref{fig5}d) illustrates a nonlinear relationship between nitrogen content and performance. Specifically, excess nitrogen incorporation adversely impacts catalytic function, whereas reduced nitrogen counts ($\rm D_N = 1, 2$) demonstrate marked activity enhancement. The SHAP dependence plot for $\rm C_N$ (Fig.~\ref{fig5}e) further refines the design criteria by identifying two distinct configuration windows (centered at 3 and spanning 5-10) where catalysts exhibit heightened qualification probabilities. By integrating these parameter-specific insights with considerations of the intrinsic correlation between features $\rm D_N$ and $\rm C_N$, we identified TM-V-1N2, TM-V-2Nn ($\rm n = 2, 3, 4$) ($\rm N_V = 4, 5$) as promising catalysts with exceptional performance.

\renewcommand{\floatpagefraction}{.9}
\begin{figure*}[t]
    \centering
    \includegraphics[width=0.75\textwidth]{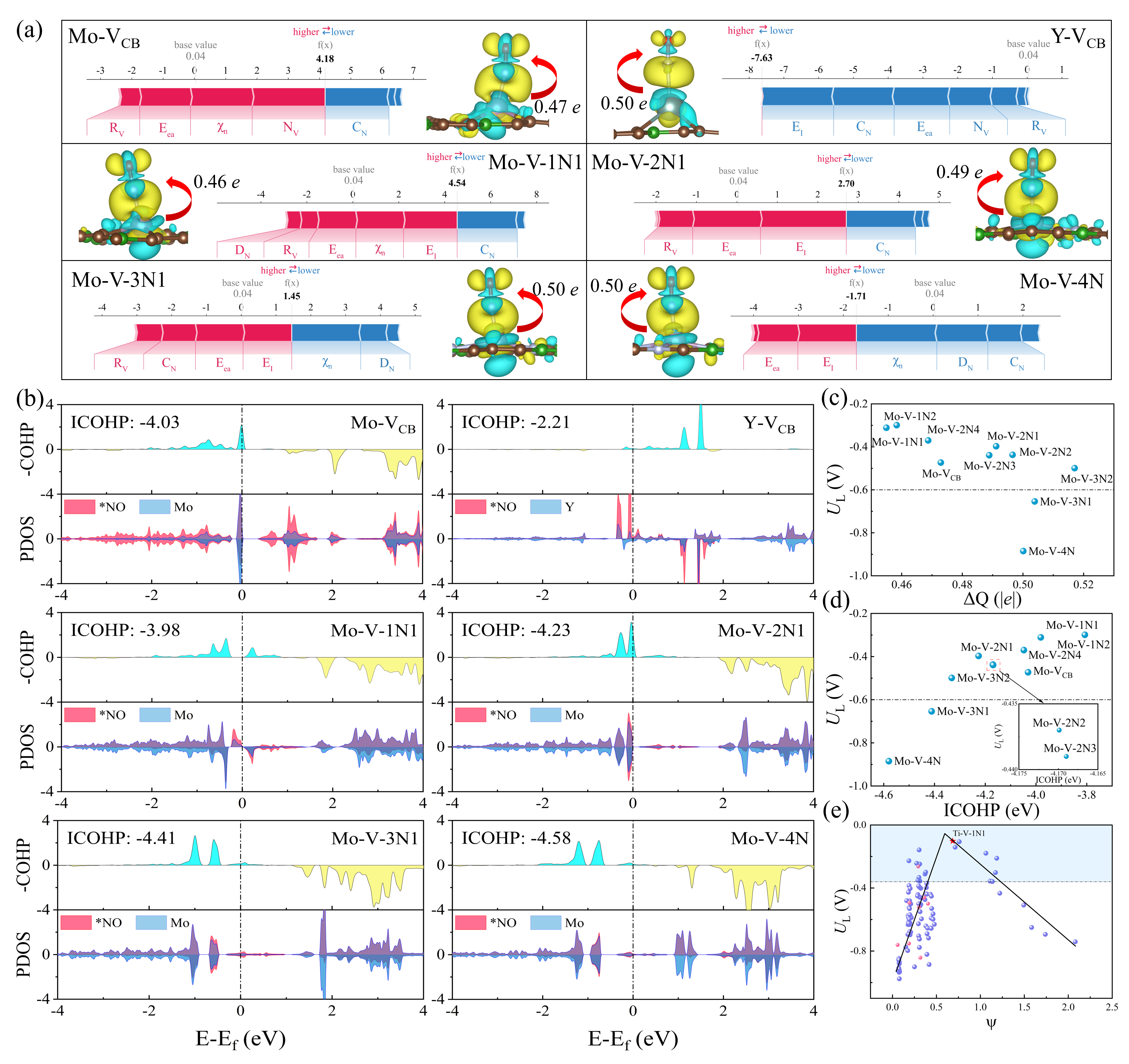}
    \caption{(a) The SHAP force analysis for Mo-$\rm V_{CB}$, Y-$\rm V_{CB}$, Mo-V-xN1 (x$=$1-3) and Mo-V-4N, where feature contributions to catalytic performance are color-mapped: positive (red) and negative (blue) impacts. Sample-specific SHAP values are displayed as bold numerals (baseline: $0.04$). Corresponding charge density difference plots (isosurface: 0.003 \textit{e} $\rm bohr^{-3}$) are inset, depicting charge accumulation (yellow) and depletion (cyan). (b) Electronic structure analysis through COHP, ICOHP, and PDOS for $\rm ^*NO$ on Mo-$\rm V_{CB}$, Y-$\rm V_{CB}$, Mo-V-xN1 and Mo-V-4N systems. (c-d) Correlation analysis of limiting potential ($U_{\rm L}$) for Mo-based catalysts (excluding Mo-$\rm V_{CC}$) versus (c) charge transfer ($\Delta {\rm Q}$) and (d) ICOHP values. (e) Volcanic plot between $U_{\rm L}$ and $\rm \psi$ (the threshold for the blue area is $U_{\rm L}$\textless\ $-$0.36 V; red points represent the 26 structures not included in the original dataset).
}
    \label{fig6}
\end{figure*}

\subsection{SHAP-aided Insight into Catalytic Activity}

To explore the electronic origins underlying catalytic performance variations, we conducted a comparative analysis focusing on the most influential feature, $\rm N_V$, leveraging SHAP force plots for two representative catalysts: Mo-$\rm V_{CB}$ ($\rm N_V = 6$) and Y-$\rm V_{CB}$ ($\rm N_V = 3$), as shown in Fig.~\ref{fig6}a (upper panel). The corresponding SHAP values of 4.18 and $-$7.63 for Mo-$\rm V_{CB}$ and Y-$\rm V_{CB}$, respectively, signify a strong positive contribution to catalytic performance in the former and a detrimental effect in the latter, aligning well with their classification as qualified and unqualified. Given the central role of $\rm ^*NO$ adsorption in the reaction pathway, partial density of states (PDOS) and crystal orbital Hamiltonian population (COHP) analyses were conducted for these two catalysts. Mo-$\rm V_{CB}$ displays significantly enhanced orbital hybridization between TM center and $\rm ^*NO$ near the Fermi level, along with a more negative integrated COHP value (ICOHP) reflecting stronger chemical bonding. In contrast, the weak orbital overlap in Y-$\rm V_{CB}$ leads to poor $\rm ^*NO$ activation and a substantially higher energy barrier for the initial hydrogenation step ($\Delta G_{\rm ^*NO\rightarrow {^*NOH/{^*NHO}}} = 1.09$\ eV), thereby suppressing its catalytic activity. These electronic observations strongly support the trends revealed by SHAP analysis, highlighting the importance of sufficient $d\mbox{-}{\rm electron}$ availability ($\rm N_V$) in promoting $\rm ^*NO$ activation and classifying low-$\rm N_V$ candidates as less active.

To evaluate the impact of $\rm C_N$, which derives from $\rm D_N$ feature, we systematically analyzed a Mo-centered catalyst series: Mo-$\rm V_{CB}$ ($\rm D_N =$ 0), Mo-V-1N1 ($\rm D_N =$ 1), Mo-V-2N1 ($\rm D_N =$ 2), Mo-V-3N1 ($\rm D_N =$ 3), and Mo-V-4N ($\rm D_N =$ 4). Among these, Mo-$\rm V_{CB}$, Mo-V-1N1, and Mo-V-2N1 were identified as qualified. SHAP force plots(Fig.~\ref{fig6}a, lower panel) revealed that $\rm D_N = 0$ made negligible contributions, a marked positive effect emerged at $\rm D_N = 1$, which then diminished at $\rm D_N = 2$ and transitioned to pronounced negative contributions at higher doping levels ($\rm D_N = 3$ and 4). PDOS and COHP analyses (Fig.~\ref{fig6}b) indicated a non-linear trend in charge transfer from the TM center to $\rm ^*NO$, which initially decreased and increased with increasing $\rm D_N$. Correspondingly, the ICOHP values exhibited a similar pattern, shifting toward more negative values with higher $\rm D_N$, suggesting enhanced $\rm ^*NO$ binding strength. Notably, the evolution in SHAP contributions, ICOHP, and charge transfer ($\Delta {\rm Q}$) consistently reflect the adverse effect of overly strong $\rm ^*NO$ adsorption on catalytic activity. Specifically, Mo-V-1N1 featured the weakest $\rm ^*NO$ binding ($\rm ICOHP=-3.98$) and the lowest energy barrier for the $\rm ^*NO\rightarrow {^*NOH/^*NHO}$ step ($\Delta G = 0.17$\ eV), but encountered a comparatively high barrier for the final $\rm ^*NH_2\rightarrow {^*NH_3}$ hydrogenation ($\Delta G = 0.31$\ eV). On the contrary, Mo-V-3N1 with stronger $\rm ^*NO$ adsorption ($\rm ICOHP=-4.41$), enabled easier hydrogenation of $\rm ^*NH_2\rightarrow {^*NH_3}$ ($\Delta G= 0.12$\ eV) but suffered elevated barriers for the initial step ($\Delta G=0.66$ eV). These trends accord with the Sabatier principle, which favors moderate adsorption strengths for optimal catalytic efficiency.\textsuperscript{\cite{hu2021sabatier}} Collectively, these findings suggest that moderate $\rm C_N$ coordination, which can be achieved through controlled nitrogen doping, strikes a balance between $\rm ^*NO$ activation and the final hydrogenation steps, thereby enhancing overall performance.

To further elucidate the effect of $\rm C_N$, we extended our analysis to Zr-V-2Nn ($\rm n = 1\mbox{-}4$) catalysts. Despite having identical composition and nitrogen substitution levels, considerable variations in the SHAP force plots, COHP, and PDOS were observed (Fig. S6). These differences stem from the specific nitrogen doping sites, which altered the degree of orbital hybridization between the TM and $\rm ^*NO$ near the Fermi level. This underscores the critical influence of the local atomic environment, including second coordination shells, on catalytic behavior and highlights the need for atom-level design precision in engineering active sites. A quantitative correlation between $\Delta {\rm Q}$ and ICOHP for $\rm ^*NO$ adsorption across Mo-based catalysts (excluding Mo-$\rm V_{CC}$) is shown in Fig.~\ref{fig6}c and 6d. As illustrated in Fig.~\ref{fig6}c, samples like Mo-V-4N and Mo-V-3N1, though unqualified, displayed moderate charge transfer. Interestingly, both excessively high and low $\Delta {\rm Q}$ values correspond to improved catalytic activity, implying that optimal performance resides in balanced electronic states. Fig.~\ref{fig6}d further confirms that less negative (or positive) ICOHP values tend to correlate with more favorable selectivity and activity. Taken together, these results demonstrate that both over-binding and under-binding of $\rm ^*NO$ may impair catalytic efficacy, and a balanced charge distribution is key.

To construct a more predictive structure–performance descriptor, we incorporated features from 107 candidates screened in Stage III with $U_{\rm L}$ \textless \ $\rm -1.0$ V (Table S1). Feature importance analysis (Fig.~\ref{fig5}a) indicates that catalytic activity is dominated by the properties of the TM center and its immediate coordination environment. We formulated normalized terms reflecting $d\mbox{-}{\rm electron}$ count ($\rm S_d = (N_V-2)/N_{d\_max} \ with \ N_{d\_max}=10$), electronegativity ($\rm S_{\chi _{TM}}=\chi _{TM}/\chi _{TM\_max}\ with\ \chi _{TM\_max}=5.77$), and van der Waals radius ($\rm R_V$),\textsuperscript{\cite{ren2025single}} and defined the TM contribution term as $\rm S_d \times S_{\chi_{TM}}/R_V$ to encapsulate the combined electronic, chemical, and geometric effects. For the local coordination environment, we normalized the electronegativity of nearest neighbors as $\rm S_{\chi_n}=\chi_n/\chi_{n\_max}\ (\chi_{n\_max}=34.7)$. 

When plotting $U_{\rm L}$ against $\rm S_d \times S_{\chi_{TM}} \times S_{\chi_n}/R_V$ (Fig. S7), scattered points appeared in the mid-range. Analysis revealed these deviations arose from $\rm ^*NOH/^*NHO$ intermediates formed after the initial hydrogenation of $\rm ^*NO$. To account for this, we introduced $\rm \theta$, defined as the deviation of the O–N–H bond angle in $\rm ^*NOH/^*NHO$ from 90° (Fig. S8). The descriptor was expressed as:

\begin{equation*}
    \rm \psi=\frac{S_d \times S_{\chi_{TM}} \times S_{\chi_n} \times \theta}{R_V}
\end{equation*}

Plotting $\rm \psi$ versus $U_{\rm L}$ (Fig.~\ref{fig6}e) yielded a volcano-shaped curve with strong linearity. Moreover, calculating the $\psi$ values for the 26 structures excluded from the original dataset revealed that Ti-V-1N1 lies near the volcano apex, suggesting its potential as the most active catalyst. These 26 structures were then subjected to the same DFT calculations and high-throughput screening, with $U_{\rm L}$ and $\psi$ values of structures that passed the first three steps of high-throughput screening and $\Delta G_{\rm max}$ less than 1.0 eV included in Fig.~\ref{fig6}e as red data points. Remarkably, these newly evaluated structures also aligned well with the original volcano trend, demonstrating the predictive capability of $\psi$ in estimating $\rm NO_3RR$ catalytic activity for both TM-$\rm V_{CC/CB}$ and heteroatom-doped systems. The best-performing Ti-V-1N1 catalyst remains positioned closest to the apex, further confirming $\psi$ as a valid and effective descriptor for catalytic performance evaluation. Catalysts with $U_{\rm L}$ \textless $\rm -0.36$ V, located in the blue-highlighted region, are designated as highly active and will be comprehensively assessed in the following section. Notably, an unoccupied region persists at the volcano peak. Based on this descriptor, we hypothesize that further optimization of heteroatom type, quantity, and spatial distribution, in conjunction with targeted TM selection, may drive $\psi$ closer to the apex, enabling even higher catalytic activity. Note that, most of the present high-performance catalytic systems position close to the peak of the volcano-shaped curve prefer relatively low spin states, which may facilitate the modest NO activation according to the spin selection rule.\textsuperscript{\cite{wigner1928struktur,moore1973investigation,li2016interplay,zhang2022synergetic,zhang2025highly}} Importantly, $\psi$ is primarily derived from intrinsic material properties and requires minimal DFT input, only for $\theta$, enabling efficient evaluation.

\renewcommand{\floatpagefraction}{.9}
\begin{figure*}[t]
    \centering
    \includegraphics[width=0.8\textwidth]{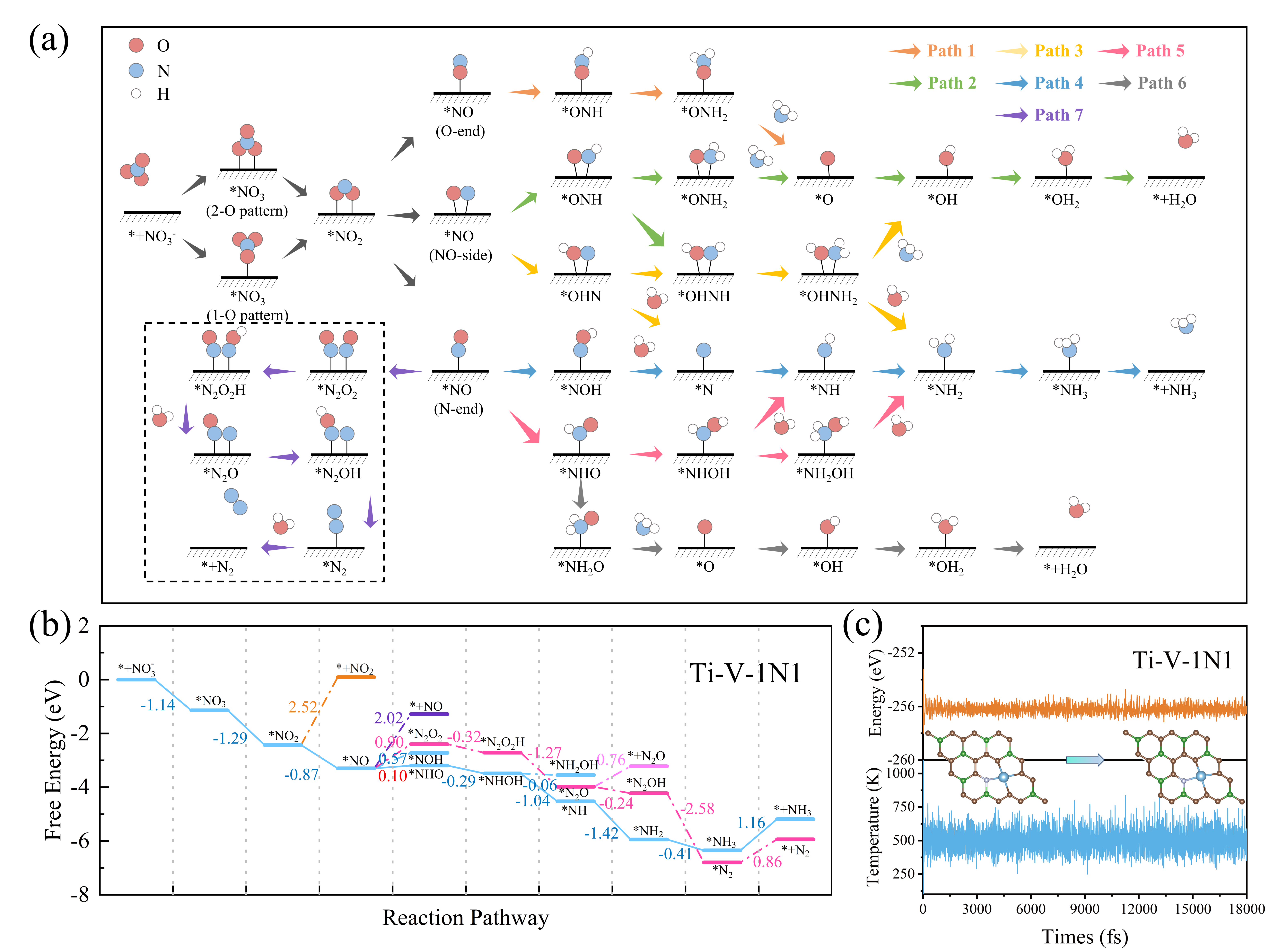}
    \caption{(a) Proposed mechanism pathways for $\rm NO_3RR$ with  associated transition states. Thermodynamic and kinetic analysis of catalytic systems: (b) Gibbs free energy profiles and (c) AIMD simulations at 500 K for Ti-V-1N1, with structural evolution illustrated through top-view snapshots of initial and final states.}
    \label{fig7}
\end{figure*}

\subsection{Optimal Reaction Pathway Screening and Stability Analysis}
Through systematic evaluation of catalytic performance metrics, a subset of 16 catalysts was identified as exhibiting superior activity based on the criterion of $\Delta G_{\rm max}$ \textless \ 0.36 eV for the two critical hydrogenation steps. This selection process, however, must be contextualized within the inherent complexity of the $\rm NO_3RR$ mechanism, which comprises a multi-step hydrogenation network generating diverse reaction pathways. To address this systems-level challenge, we constructed comprehensive reaction pathway diagrams (Fig.~\ref{fig7}a), explicitly incorporating the competitive byproduct formation channel (Path 7) as previously reported.\textsuperscript{\cite{wang2020enhanced, jia2020boosting}} The diagrammatic analysis revealed that the dominant $\rm NH_3$ production pathways are fundamentally governed by the adsorption configuration of the $\rm ^*NO$ intermediate. Notably, N-end coordination emerges as the predominant mode across the 117 candidates pre-screened via high-throughput selection and the subsequently validated supplementary structures (Fig.~\ref{fig1}b) . This observation justified our focused investigation of Path 4-7 as the most kinetically viable reaction channels (Fig.~\ref{fig7}a).

To dissect the governing principles of these reaction mechanisms, we constructed Gibbs free energy profiles for the 16 leading catalysts (Fig.~\ref{fig7}b, Fig. S9, Fig. S10). These energetic landscapes expose decisive control points where byproducts formation is thermodynamically impeded. Specifically, PDS consistently manifest during either $\rm ^*NO \rightarrow {^*NOH/^*NHO}$ hydrogenation or $\rm ^*NH_2 \rightarrow {^*NH_3}$ formation. Exemplary analysis of Ti-V-1N1 (Fig.~\ref{fig7}b) demonstrates that direct desorption of $\rm NO_2/NO$ intermediates and $\rm ^*N_2O_2$-mediated $\rm N_2$ evolution are energetically disfavored ($\Delta G$ \textgreater \ 0.36 eV). Instead, the system progresses through Path 5 where $\rm ^*NHOH$ dehydrates to *NH ($\Delta G \ {\rm = 0.10\ eV}$), followed by Path 4 for $\rm NH_3$ production under a limiting potential of $\rm -0.10$ V, is predicted to surpass NbV-$\rm N_4$ ($U_{\rm L}=-0.20$ V), Cu/Ni-NC ($U_{\rm L}=-0.37$ V), Cu-$\rm N_3$-tube ($U_{L}=-0.11$ V), and most reported $\rm NO_3RR$ electrocatalysts.\textsuperscript{\cite{lv2022tunable, wang2023n, liang2025rational}}

To the end, the operational viability of these catalysts was further corroborated through AIMD simulations at 500 K. After 18 ps of thermal annealing, all 16 systems maintained structural integrity, as evidenced by minimal energy fluctuations (\textless \ $\rm 2.5\ eV$) and preservation of coordination geometries relative to their initial states (Fig.~\ref{fig7}c, Fig. S9, Fig. S10). These findings demonstrate that the catalysts maintain thermodynamic stability under the simulated reaction conditions, supporting their potential for practical electrochemical ammonia synthesis.

\section{Conclusions}
In summary, we have systematically uncovered the key factors governing $\rm NO_3RR$ catalytic performance through the integration of IML and DFT calculations. A comprehensive dataset composed of 286 SACs, generated via high-throughput screening, was categorized into ``qualified'' and ``unqualified'' candidates. To address class imbalance, a binary XGBoost classification model was developed using an over-sampling strategy, achieving accurate predictive performance. SHAP analysis identified three dominant features, i.e., the number of valence electrons ($\rm N_V$) of the TM, nitrogen doping number ($\rm D_N$), and nitrogen coordination configuration ($\rm C_N$), as critical determinants of catalytic activity. Subsequent investigations revealed that TM-V-1N2 and TM-V-2Nn ($\rm n = 2, 3, 4$) families, with $\rm N_V=4\ and\ 5$, displayed exceptional catalytic potential. Electronic structure analysis further confirmed the intrinsic activity of these candidates, in agreement with the high-throughput screening results. Moreover, building upon these findings, we proposed a comprehensive descriptor, which integrates key physicochemical features and the O-N-H bond angle of crucial reaction intermediates ($\rm \theta$). The $\psi$ descriptor exhibits a well-defined volcano-type relationship with the limiting potential ($U_{\rm L}$), underscoring the pivotal role of transition metal coordination environments in catalytic optimization. By applying this framework, we identified 16 high-performing, non-precious metal $\rm NO_3RR$ catalysts with $U_{\rm L}$ values all below $\rm -0.36$ V. Among them, Ti-V-1N1 is predicted to exhibit an ultralow $U_{\rm L}$ of $\rm -0.10$ V, outperforming most previously reported catalysts. This work not only delivers promising candidates for efficient nitrate reduction but also establishes a robust and general approach for the rational design of advanced electrocatalytic materials.

\begin{suppinfo}

\begin{itemize}
  \item Supporting Information.pdf: Calculation details, stability heatmap, ML fitting results, ROC curves, SHAP force plots, Gibbs free diagrams, AIMD plots, and $\psi$ and $U_{\rm L}$ values of SACs.
\end{itemize}

The authors declare no potential conflict of interests.

\end{suppinfo}

\begin{acknowledgement}

This work is supported by the National Natural Science Foundation of China (Grant No. 12374061, U23A2072, 12404274), Zhejiang Provincial Natural Science Foundation of China (Grant No. LD24F040001, LQN25A040020), and the KC Wong Magna Foundation in Ningbo University.

\end{acknowledgement}

\end{document}